\documentstyle [12pt,aps,amsfonts] {revtex} 
\input epsf
\newcommand{\beq}{\begin{equation}}
\newcommand{\eeq}{\end{equation}}
\newcommand{\beqs}{\begin{eqnarray}}
\newcommand{\eeqs}{\end{eqnarray}}
\newcommand{\lsim}{\mathrel{\raisebox{-.6ex}{$\stackrel{\textstyle<}{\sim}$}}}

\begin{document}
\tighten
\draft

\baselineskip 6.0mm

\title{Weak and Electromagnetic Nuclear Decay Signatures for Neutrino 
Reactions in SuperKamiokande} 

\vspace{8mm}

\author{
Shmuel Nussinov$^{(a,b)}$ \thanks{email: nussinov@post.tau.ac.il} \and
Robert Shrock$^{(b)}$ \thanks{email: robert.shrock@sunysb.edu}}

\vspace{6mm}

\address{(a) \ Sackler Faculty of Science \\
Tel Aviv University \\
Tel Aviv, Israel} 

\address{(b) \ C. N. Yang Institute for Theoretical Physics \\
State University of New York \\
Stony Brook, N. Y. 11794, USA}

\maketitle

\vspace{10mm}

\begin{abstract}

We suggest the study of events in the SuperKamiokande neutrino data due to
charged- and neutral-current neutrino reactions followed by weak and/or
electromagnetic decays of struck nuclei and fragments thereof.  This study
could improve the prospects of obtaining evidence for $\tau$ production from
$\nu_\mu \to \nu_\tau$ oscillations and could augment the data sample used to
disfavor $\nu_\mu \to \nu_{sterile}$ oscillations. 

\end{abstract}

\vspace{16mm}

\pagestyle{empty}
\newpage

\pagestyle{plain}
\pagenumbering{arabic}
\renewcommand{\thefootnote}{\arabic{footnote}}
\setcounter{footnote}{0}

The very large SuperKamiokande (SK) underground water Cherenkov detector has
contributed greatly to our knowledge of solar and atmospheric neutrino physics
\cite{sol1,sol,sk,nonus}, as well as yielding stringent limits on nucleon decay
\cite{skpdec}.  The atmospheric data has, with high statistics, provided
evidence for neutrino oscillations and hence neutrino masses and lepton mixing;
the best fit is to $\nu_\mu \to \nu_\tau$, with $\Delta
m^2_{atm}\sim 3.5 \times 10^{-3}$ eV$ ^2$ and maximal mixing, $\sin^2
2\theta_{atm} = 1$ \cite{sk}.  These values are also consistent with data from
the Soudan-2 \cite{soudan2} and MACRO \cite{macro} experiments, as well as the
K2K long baseline experiment \cite{k2k}.  SK has also obtained a large sample
of solar neutrino events, yielding a confirmation of a factor of 2 deficit
relative to calculations, measurement of the energy distribution, and tests of 
day-night asymmetry and seasonal variation. 

The main neutrino reactions that have been analyzed in SK include, for solar
neutrinos, $\nu_e e$ elastic scattering and, for atmospheric neutrinos,
quasi-elastic charged-current (CC) scattering, $\nu_\ell + n \to \ell^- + p$,
$\bar\nu_\ell + p \to \ell^+ + n$, where $\ell=e,\mu$ \cite{sk}, and
neutrino-induced single and multi-pion production \cite{nonus}. These all
involve prompt particle production.  However, because the primary neutrino
reactions can lead to excited levels of nuclei or to nuclei with unstable
ground states (g.s.), $\gamma$ and delayed $\beta$ and $\beta\gamma$ decays can
occur in conjunction with these primary neutrino reactions and the hadronic
cascade which may accompany them.  This is particularly true of atmospheric
neutrinos because their energies extend to $\sim$ GeV, as contrasted with the
14.5 MeV upper end of the solar $^8$B $\nu_e$'s.  When the released energy from
these decays exceeds about 5-6 MeV, they should be detectable.  Indeed, SK has
successfully measured scattered electrons with energies of 5 MeV in their solar
neutrino data sample, despite the backgrounds that become increasingly severe
at low energies \cite{sol1,sol}.  Here we suggest the study of these events and
analyze some of their signatures.  In 1144 days of running, SK has obtained
9178 atmospheric neutrino events that are fully contained (FC) in the 22.5 kton
fiducial part of the 50 kton total volume \cite{sk,nonus}; from this it has
been estimated that SK should have approximately 70 events involving
charged-current $\tau^\pm$ production by $\nu_\tau$ and $\bar\nu_\tau$
originating from the inferred $\nu_\mu \to \nu_\tau$ and $\bar\nu_\mu \to
\bar\nu_\tau$ oscillations \cite{sk}.  The additional information that the
$\gamma$, $\beta$ and $\beta\gamma$ decays provide could improve the
understanding of both single-ring and multi-ring events, helping the efforts to
pick out these $\tau^\pm$ events.

Many of the primary charged- and neutral-current (CC and NC) neutrino
interactions, as well as secondary interactions of pions, nucleons, and
possibly larger nuclear fragments, occur on the $^{16}$O nuclei.  These
interactions can involve charge exchange, further emission of one or more
nucleons, breakups into fragments, and excitations that promptly decay via
nucleon emission.  In many of these cases, these process leave unstable nuclei
at the locations (vertices) of the original neutrino reactions.  These nuclei
may have been excited to higher levels that decay electromagnetically to the
respective ground states promptly via photon emission. Such decays
coincide, to within the experimental time resolution of the SK detector, with
the initial event and yield monochromatic $\gamma$'s.  There can occur ground
states of nuclei $(Z,A)$ that subsequently beta decay to neighboring nuclei $(Z
\pm 1,A)$.  For several nuclei with $A \le 16$, the maximum $e^\pm$ energies,
i.e., $Q_\beta$ values, exceed the expected 5-6 MeV threshold for detection
above background.  The lifetimes of these beta decays range between $\sim 0.01$
sec and 10 sec, well beyond the time window of the original event.  Hence these
beta decays occur during a time when the original Cherenkov signals from the
primary reaction have ceased, and therefore would be not be confused with them.
The same struck oxygen nucleus will serve as the origin for the Cherenkov cone
associated with the delayed $\beta$ (and {\it a fortiori} any prompt $\gamma$)
and one or several cones in the initial neutrino event and its associated
hadronic cascade.  The beta decay times are short enough to prevent any
significant motion of the radioactive nuclei away from the location of the
primary vertex due to the momentum transfer in the primary neutrino reaction or
the motion of the water owing to its recirculation (at a rate of 50 tons/hour
\cite{ishihara}).  The recoil of the nucleus or fragment in question is of
order the Fermi momentum $p_F \sim 300$ MeV/c (larger momentum transfers
typically result in further fragmentation); hence the range of
this nucleus is much shorter than the spatial resolution with which the
location of the primary event vertex can be reconstructed.  This spatial
resolution depends on the angle with respect to the Cherenkov cone and with the
lepton energy \cite{sk}; here, to be conservative in our background estimates,
we shall take it to be 1 m.

Accidental backgrounds are negligible for the cases of decays of excited nuclei
by prompt photon emission, since these are simultaneous, to detector time
resolution, with the initial event.  Next consider beta decays.  From the 9178
fully contained atmospheric neutrino events in 1144 days, it follows that these
events occur at an average rate of 0.33 per hour.  The key discriminant here is
that the Cherenkov cones or showers associated with the delayed $\beta$ and/or
$\beta\gamma$ decays must originate from a volume of $\sim 1$ m$^3$ out of the
$2.25 \times 10^4$ m$^3$ fiducial volume.  Over a time window after the primary
reaction of 30 sec, say, accidental background events of this type should
therefore be negligible. A particular background arises from decays of nuclei
that are spallation products of collisions of through-going muons.  These were
also a background for the solar neutrino signal and were effectively removed by
rejecting events in which a beta emerges from a cylindrical region around the
path of the through-going muon during a time from msec to O(10) sec after this
muon traverses the detector \cite{sol1,sol,cy}.  The same method could be used
here.  An additional check on backgrounds is obtained from the SK data on
$e$-type single-ring solar neutrino events as a function of angle with respect
to the sun, $\cos\theta_{sun}$ \cite{sol1}.  This data showed a background, of
$\sim 0.09$ events/day/kton/bin, isotropic in $\cos\theta_{sun}$ (and the solar
neutrino signal above it, at $\cos\theta_{sun} \simeq 1$); this number yields,
in 1 m$^3$ volume, over a 30 sec time window, a negligibly small rate for this
background. Having outlined the general idea, we proceed to specific cases.

As a first example for atmospheric neutrinos, consider the reaction
\beq
\bar\nu_\ell + ^{16}{\rm O} \to \ell^+ + ^{16}{\rm N}
\label{nubaroxn16}
\eeq 
where $\ell=e$ or $\mu$. The g.s. of the $^{16}$N nucleus beta decays,
with half-life $t_{1/2} = 7.1$ sec, to various levels of $^{16}$O, primarily
the g.s., with branching ratio $BR = 26$ \% and maximum $e^-$ energy
$Q_\beta=10.42$ MeV, and the $3^-$ state, with $BR=68$ \% and $Q_\beta = 4.29$
MeV, which then decays promptly (in 18 psec) to the g.s. via emission of a
monochromatic 6.13 MeV photon (this and other nuclear properties are from
\cite{nucprop}).  The selection criteria for these events would be (i) fully
contained (FC) within the 22.5 kton fiducial volume, with (ii) a Cherenkov ring
from the initial $\ell^+$, (iii) for $\ell^+=\mu^+$, a second, $e$-type ring
from the $\mu^+$ decay, (iv) in a timing window of $0 \le t \le 30$ sec after
the initial signal the appearance of an $e^-$ with a reasonable fraction of
10.4 MeV, from the direct decay, or a 6.13 MeV $\gamma$ from the branched
decay, satisfying the requirement that (v) the $e^-$ Cherenkov showering cone
or the shower from the $\gamma$ extrapolate back to the vertex reconstructed
from the initial $\ell^+$ Cherenkov signal.  Note that SK has achieved an 
accuracy of better than 1 \% in the energy scale and about 15 \% in 
$\Delta E/E$ for 10 MeV electrons in its solar neutrino data \cite{sol1,sol}. 
Some relevant cross section calculations for the primary atmospheric neutrino 
reactions are \cite{sigma},\cite{nucrev}.  For a rough estimate of the 
fraction of the reactions (\ref{nubaroxn16}) that yield 
$^{16}{\rm N}_{g.s.}$, we may refer to the CRPA (continuous random phase
approximation) calculations of analogous CC neutrino reactions on 
$^{12}{\rm C}$ \cite{carboncal1} 
(see also \cite{carboncal2},\cite{carboncal3}) performed to compare with LSND 
data \cite{lsndcarbon}. It was found that reactions
yielding nuclides close to the valley of beta stability occurred with sizable
fractions of the inclusive yield; e.g., the flux-averaged cross sections for 
$^{12}{\rm C}(\bar\nu_\mu,\mu^+)^{12}{\rm B}_{g.s.}$ and 
$^{12}{\rm C}(\bar\nu_\mu,\mu^+n)^{11}{\rm B}_{g.s.}$ comprised, respectively,
6 \% and of 34 \% of the inclusive process $^{12}{\rm C}(\bar\nu_\mu,\mu^+)X$. 
 From these, we estimate that the atmospheric neutrino flux-averaged
cross section for $^{16}{\rm O}(\bar\nu_\ell,\ell^+)^{16}{\rm N}_{g.s.}$ 
could be $\sim O(10)$ \% of $^{16}{\rm O}(\bar\nu_\ell,\ell^+)X$.  
Using this ratio and the flux-averaged $\sigma$'s for 
$\bar\nu_\ell p \to \ell^+ n$ and $^{16}{\rm O}(\bar\nu_\ell,\ell^+)X$ from 
\cite{sigma}, we estimate that of the 3107 $e$-like and 2988 $\mu$-like
single-ring FC events in the 1144 day SK
data, there could be $O(10^2)$ events of type (\ref{nubaroxn16}) leading to 
$^{16}{\rm N}_{g.s.}$, of which about 40 \% would have $E_e > 6$ MeV.

Another pathway for producing $^{16}$N is a regular fully contained event
$\nu_\mu n \to \mu^- p$ where the $\mu^-$ is captured on another oxygen nucleus
via $^{16}$O$(\mu^-,\nu_\mu)^{16}$N, followed by the beta decay of the
$^{16}$N.  Such an event can be distinguished from those which we propose to
study because for typical $E_\mu$, the $\mu^-$ travels sufficiently far from
the primary event vertex that criterion (v) above would not be satisfied; for
example, the $\mu^-$ range $R \sim 2$ m for $p_\mu=500$ MeV/c.  We note that SK
has been able to measure beta decays of $^{16}$N nuclei produced both
artificially, by the $^{16}$O$(n,p)^{16}$N reaction, and naturally, by stopping
$\mu^-$'s captured on $^{16}$O \cite{n16}.  In 1004 days of data, a rate of
$11.4 \pm 0.2$ events/day from the $^{16}$O$(\mu^-,\nu_\mu)^{16}$N reaction and
subsequent $^{16}$N beta decay in an inner 11.5 kton fiducial volume was
measured, in agreement with a prediction of $11.9 \pm 1.0$ events/day. 
 Note that SK, in its search for the nucleon
decay $p \to \bar\nu K^+$, used as an event criterion the observation of a
6.32 MeV photon from a resultant excited $^{15}$N nucleus \cite{skpdec}.

Neutrino reactions can lead to the fragmentation of the
$^{16}$O nucleus with emission of protons, neutrons, and higher-$A$ fragments. 
For example, consider 
\beq
\bar \nu_\ell + ^{16}{\rm O} \to \ell^+ + p + ^{15}{\rm C} \ . 
\label{nuoxc15p}
\eeq 
The g.s. of $^{15}$C beta decays, with $t_{1/2}=2.4$ sec, to (i) the
g.s. of $^{15}$N, with $Q_\beta=9.77$ MeV, $BR=32$ \%, and (ii) the 5.30
MeV level, with $Q_\beta=4.47$ MeV, $BR=68$ \%, followed promptly by
decay to the g.s. with emission of a 5.30 MeV photon.  Although the branched
decay would be difficult to detect, the energy of the $e^-$ could be large
enough to detect. 

Higher-$A$ fragmentation reactions include 
\beq
\bar\nu_\ell + ^{16}{\rm O} \to \ell^+ + ^{3}{\rm He} + ^{13}{\rm B} \ . 
\label{nubaroxbhe3}
\eeq
The $^{13}$B nucleus beta decays with $t_{1/2}=17$ msec, mainly ($BR=93$ \%) 
to the g.s. of $^{13}$C, emitting an $e^-$ with $Q_\beta=13.4$ MeV. 
A second example is 
\beq
\bar\nu_\ell + ^{16}{\rm O} \to \ell^+ +  ^{4}{\rm He} + ^{12}{\rm B} \ . 
\label{nubaroxbhe}
\eeq
Here the $^{12}$B nucleus beta decays with $t_{1/2}=20$ msec to several 
levels of $^{12}$C, mainly ($BR=97$ \%) to the g.s. with 
$Q_\beta=13.4$ MeV.  A third case is 
\beq
\nu_\ell + ^{16}{\rm O} \to \ell^- + ^{4}{\rm He} + ^{12}{\rm N}
\label{nuoxnalpha}
\eeq
followed by the beta decay $^{12}{\rm N} \to e^+ + \nu_e + ^{12}$C with 
$t_{1/2}=11$ msec, mainly ($BR=94$ \%) the g.s. with $Q_\beta=16.3$ MeV. 
For a rough estimate of the cross sections for (\ref{nubaroxbhe}) and 
(\ref{nuoxnalpha}), we note that for the similar CC reactions on $^{12}$C, the
CRPA calculation of \cite{carboncal1} found that the flux-averaged $\sigma$'s
for $^{12}{\rm C}(\nu_\mu,\mu^- \alpha)^{8}{\rm B}$, 
$^{12}{\rm C}(\nu_\mu,\mu^- p \alpha)^{7}{\rm Be}$, 
$^{12}{\rm C}(\bar\nu_\mu,\mu^+ \alpha)^{8}{\rm Li}$, and 
$^{12}{\rm C}(\bar\nu_\mu,\mu^+ n \alpha)^{7}{\rm Li}$ comprised 
7.5 \%, 18 \%, 7.5 \%, and 13 \% of the respective inclusive reactions 
$^{12}{\rm C}(\nu_\mu,\mu^-)X$ and $^{12}{\rm C}(\bar\nu_\mu,\mu^+)X$.
This suggests that, if comparable rates apply for (\ref{nubaroxbhe}) and 
(\ref{nuoxnalpha}), the current SK data could contain $O(10^2)$ events from
these reactions, summed over $e$ and $\mu$ type (anti)neutrinos, and 60 \% (74
\%) of the resultant $^{12}$B ($^{12}$N) beta decays would yield $e^\mp$
energies above 6 MeV. 

There are also events of the form $^{16}{\rm O}(\bar\nu_\ell,\ell^+)^{16}{\rm
N}^*$ in which the final state $^{16}$N is in an excited state.  Except for
low-lying levels with excitation energies less than 0.4 MeV above the g.s.,
these decay promptly with neutron emission. 

Among neutral-current reactions one has, for atmospheric neutrinos, 
\beq
(\nu_\ell,\bar\nu_\ell) + ^{16}{\rm O} \to (\nu_\ell,\bar\nu_\ell) + \pi^+ 
+ \ ^{16}{\rm N} \ . 
\label{ncn16pip}
\eeq
The selection criteria for events of this type would be (i) FC, (ii) Cherenkov
ring from the $\pi^+$, (iii) observation of the $\beta$ or $\beta\gamma$ decay
of the $^{16}$N, as described after (\ref{nubaroxn16}).  One could also look
for the reactions 
\beq
(\nu_\ell,\bar\nu_\ell) + ^{16}{\rm O} \to (\nu_\ell,\bar\nu_\ell) + 
\pi^0  + \ ^{16}{\rm O}^*  
\label{nuoxnc}
\eeq 
which populate excited states of the $^{16}$O nucleus.  Although the
lowest excited state, at 6.05 MeV, is a $0^+$ state and hence cannot decay to
the $0^+$ g.s. with the emission of a single photon, the $1^-$ state at 7.12
MeV decays promptly (in 8 fs) via an E1 transition, and one could try to detect
the resultant photon.  The selection criteria would be (i) FC, (ii) two photons
with invariant mass $M_{\gamma\gamma}=m_{\pi^0}$, (iii) a third photon from the
primary event vertex, to within experimental uncertainty, with $E_\gamma=7.12$
MeV.  Similar comments apply for electromagnetic decays from other excited
states. This event selection could augment the sample of neutral-current (NC) 
events that might be used to disfavor further the $\nu_\mu \to \nu_{sterile}$ 
transition, strengthening the results of \cite{nonus}.  Although there is 
some uncertainty in theoretical model calculations of cross sections for NC 
pion production, and hence also of (\ref{ncn16pip}) and (\ref{nuoxnc}), 
direct measurements of neutrino-induced pion production in the K2K near 
detector \cite{k2k} should ameliorate this problem. 

Nucleons and pions emitted in primary reactions can lead to secondary ones, via
$n + (Z,A) \to p + (Z-1,A)$, $p + (Z,A) \to n + (Z+1,A)$, and 
$\pi^\pm + (Z,A) \to \pi^0 + (Z \pm 1,A)$ charge-exchange processes.  Other 
possibilities include $(p,d)$, $(n,d)$ reactions, etc. 
The probability of excitation of any specific nuclear state in the primary
(anti)neutrino-oxygen or secondary (pion or nucleon)-oxygen collisions may be
low, and SK does not have 100 \% efficiency for detecting the resulting,
relatively low-energy $\gamma$, $\beta$, or $\beta\gamma$ signal.  However, we
are considering the {\it inclusive} rate for producing the (g.s. or excited
state) of a nucleus with the requisite $\gamma$, $\beta$, or $\beta\gamma$
decay.  Just as the many exclusive neutrino reactions sum up to yield a
linearly rising inclusive neutrino cross section for energies beyond $\sim 1$
GeV, so also, the sum of the specific exclusive reactions considered here,
leading to nuclear states that decay by $\gamma$, $\beta$, or $\beta\gamma$
emission, should be a significant fraction of the total. We note further that
in a generic high-energy neutrino event or when a $\tau^\pm$ is produced and
decays, there is substantial probability for one or more pions to be emitted.
There can thus be a resultant hadronic cascade involving collisions with
several oxygen nuclei, leading to nuclear states that decay via $\gamma$,
$\beta$, or $\beta\gamma$ decays.  Indeed, this tends to happen precisely in
the complex, multi-ring events that are difficult to analyze and for which even
partial additional information locating a particular emission vertex would be
useful.  As noted, the study of these reactions could thus aid in extracting
$\tau$'s in the atmospheric data.  

Similar comments can be made about neutrino reactions on $^{12}$C, the nucleus
(in addition to H) in a liquid scintillator.  Indeed, as part of the 
selection criteria for $^{12}$C$(\nu_\ell,\ell^-)^{12}{\rm N}_{g.s.}$ events,
the LSND experiment included detection of the $e^+$ from the beta decay
$^{12}{\rm N} \to e^+ + \nu_e + ^{12}{\rm C}$ ($t_{1/2}=11$ msec). 
The mini-BooNE experiment at Fermilab will also measure the CC $\nu_\mu$
reaction on carbon. 

In the solar neutrino data, there is the NC reaction 
\beq
\nu_e + ^{16}{\rm O} \to \nu_e + \ ^{16}{\rm O}^*
\label{nco16star}
\eeq 
producing an excited state of $^{16}$O.  In contrast to the case with
atmospheric neutrinos, here the transitions to the excited states are somewhat
suppressed since much of the $^8$B $\nu_e$ flux is below threshold.  The $0^+$
state at 6.05 MeV (reached via an allowed $0^+ \to 0^+$ Fermi transition) does
not decay via single photon emission.  The other low-lying excited states,
e.g. the $1^-$ level at 7.12 MeV, require higher-$L$ transitions from the
$0^+$ g.s. of $^{16}$O. Neutral current neutrino-nucleus reactions on nuclei 
as means of observing solar and supernova neutrinos have also been discussed in
\cite{raghavan,lvk96}.

In future work it would be worthwhile to carry out a unified Monte Carlo
simulation with joint particle-nuclear inputs of the numerous nuclear
production and decay pathways.  In general, complicated phenomena can be better
understood when viewed with a variety of wavelengths and/or times.  Our
discussion illustrates this \cite{dmeson}.  Other examples include SN1987A seen
via neutrinos \cite{bt} and in the optical, and gamma ray bursts seen with
X-ray and via optical afterglows.

S. N. would like to thank the Israeli Academy for Fundamental Research for a
grant.  The research of R. S. was partially supported by the U.S. National
Science Foundation under the grant PHY-97-22101.

\vfill
\eject

\end{document}